\documentclass{article}
\usepackage[a4paper,top=3cm,bottom=3.5cm,left=3.2cm,right=3.2cm]{geometry}

\usepackage[utf8]{inputenc}
\usepackage{authblk}
\usepackage{graphicx}
\usepackage{subfigure}
\usepackage{amssymb}
\usepackage{lineno}
\usepackage[english]{babel}
\usepackage{palatino, url}
\usepackage{amsmath}

\usepackage{listings}
\usepackage{graphicx}
\usepackage[rightcaption]{sidecap}
\usepackage{textcomp}
\usepackage{array} 
\usepackage{booktabs} 
\usepackage{multirow} 
\usepackage{caption} 
\usepackage{floatflt}
\usepackage{rotating} 
\usepackage{scrextend}
\usepackage{multirow}
\usepackage[numbers]{natbib}
\usepackage{soul}
\usepackage{mathtools}

\title{Efficient gPC-based quantification of probabilistic robustness\\ for systems in neuroscience}

\author[1]{Uros Sutulovic}
\author[1]{Daniele Proverbio}
\author[1]{Rami Katz}
\author[1]{Giulia Giordano}
\affil[1]{\small Department of Industrial Engineering, University of Trento, Trento 38123, IT }

\begin{document}

\maketitle

\begin{abstract}
Robustness analysis is very important in biology and neuroscience, to unravel behavioural patterns of systems that are conserved despite large parametric uncertainties. To make studies of probabilistic robustness more efficient and scalable when addressing complex models in neuroscience, we propose an alternative to computationally expensive Monte Carlo (MC) methods by introducing and analysing the generalised polynomial chaos (gPC) framework for uncertainty quantification. We consider both intrusive and non-intrusive gPC approaches, which turn out to be scalable and allow for a fast comprehensive exploration of parameter spaces. Focusing on widely used models of neural dynamics as case studies, we explore the trade-off between efficiency and accuracy of gPC methods, and we adopt the proposed methodology to investigate parametric uncertainties in models that feature multiple dynamic regimes.
\end{abstract}

\section{Introduction and Motivation} 
Systems biology \cite{alon2019introduction} employs methods from mathematics, physics and engineering to understand, predict and control biological systems at all scales. Of great relevance to such efforts is the study of robustness \cite{Kitano2007towards,MTNS2024}. Natural systems showcase an impressive complexity; yet, they manage to thrive despite large uncertainties, variability, or external perturbations. Examples include cell homeostasis \cite{Aoki2019}, multicellular coordination \cite{proverbio2020assessing}, and robust neural modulation \cite{goldman2001global}. Structural analysis \cite{blanchini2021structural} and robustness analysis \cite{Barmish1994} offer tools to systematically guarantee that a property of interest is preserved regardless of parameter values, or for all parameter values within a certain set. However, sometimes a property of interest does not hold structurally, nor robustly, but with some probability that one is interested in quantifying. 

Probabilistic robustness has been fruitfully introduced and investigated in engineering \cite{tempo2013randomized}, to characterise the likelihood of a system property to hold, given the probability distribution of model parameters, and to synthesise controllers that achieve a desired objective with a guaranteed probability; 
here we focus however on the \emph{analysis} of natural systems, and not on control synthesis.
The most commonly used method for probabilistic analysis is Monte Carlo (MC), running many simulations with sampled random variables and corresponding random process realisations
. However, MC methods suffer from poor scalability and require extensive computational power and time to gain information about even relatively simple models. Flexible and efficient alternative methods to quantify probabilistic robustness, building upon uncertainty quantification approaches \cite{owen2017comparison}, would scale up the investigation of parametric uncertainties in complex models, allow the extensive exploration of large parameter spaces, and enable wider adoption of robustness analysis even when low computational power must be used.

To address these challenges, we propose the use of generalised polynomial chaos (gPC), a spectral decomposition-based method to approximate the solution of differential equations with parametric uncertainties, where the uncertainties are aimed at capturing the degree of confidence of the modeller about the actual parameter values.
First developed in the realm of uncertainty quantification \cite{pepper2019multiscale, xiu2002wiener}, gPC also found applications in other fields, such as model predictive control for stochastic systems \cite{kim2013generalised}, to replace or accelerate MC methods. Analysing uncertain systems by means of gPC is computationally efficient \cite{fisher2011optimal} and the investigation of large parameter spaces can be drastically accelerated.

Here, we systematically analyse the performance of gPC methods in providing efficient surrogate models for complex dynamical systems in neuroscience, which provide a challenging case study since they display multiple regimes (i.e., different long-term behaviours of solutions) depending on the chosen parameters.
We present the theoretical background on gPC methods and assess their performance, and possible advantages over
Monte Carlo approaches, in terms of computational efficiency and accuracy. Then, we employ gPC to assess the persistence of neuronal signalling regimes subject to parametric uncertainties in three widely used models for neural dynamics, representative of different classes: the Hindmarsh-Rose model (semi-phenomenological model for single-neuron dynamics), the Jansen-Rit model (mechanistic model for multiple connected neurons), and the Epileptor model (phenomenological model for epileptic activity in brain regions).


\section{Problem Formulation and Background}

We consider autonomous ODE systems of the form
\begin{equation}
    \dot{\mathbf{x}} = f(\mathbf{x},Z),
    \label{eq:general nonlinear uncertain system}
\end{equation}
where $\mathbf{x} \in \mathbb{R}^n$ is the state of the system and $Z \in \mathbb{R}^d$ is a parametric uncertainty. Differently from stochastic differential equations, where the dynamics are driven by a stochastic process, our source of stochasticity is induced by parametric uncertainties, i.e., we assume that $Z \colon \Omega \to \mathcal{I}_Z \subset \mathbb{R}^d$ is a vector of mutually-independent random variables defined on some probability space $(\Omega,\mathcal{F},\mathbb{P})$, where $\Omega$ denotes a sample space containing samples $\omega \in \Omega$, $\mathcal{F}\subset 2^\Omega$ is a $\sigma$-algebra and $\mathbb{P}$ is a probability measure. We assume that $Z$ is absolutely continuous with respect to the Lebesgue measure, and denote its probability density function (PDF) by $\rho_Z(z)$.  
\paragraph{Remark}
From a theoretical perspective, mutual independence of the random variables in $Z$ can be assumed without loss of generality. If the random variables in $Z$ were not mutually independent, the Rosenblatt transformation, see \cite{rosenblatt1952remarks}, could be applied to transform them into new, mutually independent random variables $\tilde{Z}$.
$\hfill\diamond$ \\

Due to the parametric uncertainty, the state in \eqref{eq:general nonlinear uncertain system} is a stochastic process $\mathbf{x}(t;Z)$. Characterizing the first moments (summary statistics) of $\mathbf{x}(t;Z)$, for example in order to approximate its characteristic function \cite{florescu2013handbook}, is of interest.

The estimation of summary statistics is often performed by a combination of explicit formulas (available only in rare cases) and randomised algorithms. Monte Carlo methods 
are widely used in computational studies, to perform random sampling for numerical simulations and extraction of summary statistics. Yet, they may be impractical when simulating large models with uncertain parameters: their sample complexity scales poorly with model dimensionality, and grows rapidly with increase in the desired accuracy \cite{tempo2013randomized}. 

To overcome such computational limitations, we need alternative surrogate models that are computationally more tractable and sufficiently accurate. For practical applications, it is also necessary to determine, at least empirically, \textit{how well} such surrogate models perform in uncertainty quantification and robustness analysis tasks. Here we propose the use of various gPC methods to estimate summary statistics. 

\subsection{Generalised polynomial chaos}


Considering the probability density function of $Z$ in \eqref{eq:general nonlinear uncertain system}, we restrict our attention to the set of functions $\psi \colon \mathcal{I}_Z\rightarrow \mathbb{R}$ that belong to the weighted $L^2$ space
\begin{equation}
    L^2_{\rho_Z}(\mathcal{I}_Z) = \left \{ \psi \, \colon \, \mathbb{E}[\psi^2] =\int_{\mathcal{I}_Z} \psi^2(z)\rho_Z(z) dz<\infty \right \}.
    \label{eq:square-integrable random variables set}
\end{equation}
A natural basis for the space $L^2_{\rho_Z}(\mathcal{I}_Z)$ is a set of multivariate polynomials $\{ \Phi_\mathbf{\alpha}(z) \}_\alpha$ satisfying the orthogonality relation
\begin{equation}
    \mathbb{E}[{\Phi_\mathbf{\alpha}\Phi_\mathbf{\beta}}] = \int_{\mathcal{I}_Z} \Phi_\mathbf{\alpha}(z)\Phi_\mathbf{\beta}(z)\rho_Z(z) dz = \gamma_\mathbf{\alpha} \delta_{\mathbf{\alpha}\mathbf{\beta}},
    \label{eq:orthogonality of polynomials}
\end{equation}
where $\gamma_\mathbf{\alpha} = \mathbb{E}[\Phi_\mathbf{\alpha}^2]>0$ is a normalizing factor, and $\delta_{\mathbf{\alpha}\mathbf{\beta}}$ is the $d$-variate Kronecker delta \cite{xiu2010numerical}. Here, $\mathbf{\alpha} = (\alpha_1,\ldots,\alpha_d) \in \mathbb{N}_0^d$ is a multi-index. The basis $\{ \Phi_\mathbf{\alpha}(z) \}_\alpha$ can be obtained from a Gram-Schmidt orthogonalisation process applied to the set of monomials $\{ \prod_{i=1}^dz_i^{\alpha_i} \}_{\alpha \in \mathbb{N}_0^d}$. Henceforth, we assume that $\gamma_{\alpha }=1$ for all $\alpha \in \mathbb{N}_0^d$, meaning that the polynomials $\left\{\Phi_{\alpha}(z) \right\}_\alpha$ are orthonormal.

Wiener \cite{wiener1938homogeneous} proved a PC expansion for a normally distributed $Z\sim N(\mathbf{0},\mathbf{I})$, where $\mathbf{I}$ is the identity matrix:
\begin{equation}
    \psi(Z) = \sum_{|\mathbf{\alpha}|=0}^\infty \hat{\psi}_\mathbf{\alpha}\Phi_\mathbf{\alpha}(Z), \quad \hat{\psi}_\mathbf{\alpha} = \mathbb{E}[\psi\Phi_\mathbf{\alpha}],
    \label{eq:PCE}
\end{equation}
where $\{ \Phi_\mathbf{\alpha}(z) \}_\alpha$ is a basis of Hermite polynomials and $|\mathbf{\alpha}|=\sum_{i=1}^d\alpha_i$. Later, Cameron and Martin \cite{cameron1947orthogonal} generalised this expansion to random variables $Z$ with an arbitrary distribution. Xiu and Karniadakis \cite{xiu2002wiener} proposed a framework that links standard random variables and their densities to the corresponding orthogonal polynomials of the Wiener-Askey table, some of which are given in Table~\ref{tab:Askey scheme}. Constructing the basis $\{ \Phi_\mathbf{\alpha}(z) \}_\alpha$ in case of arbitrarily distributed random variables is the subject of dedicated studies; for instance, we refer to the procedures described in \cite{wan2006multi}.

\begin{table}[]
\centering
\caption{\small Wiener-Askey scheme}
\begin{tabular}{l|l}
PDF of $Z$ & Basis $\{\Phi_\mathbf{\alpha}\}_\alpha$   \\ \hline
Gaussian   & Hermite polynomials \\ 
Uniform   & Legendre polynomials \\ 
Beta       & Jacobi polynomials   \\ 
Gamma      & Laguerre polynomials \\ 
\end{tabular}
\label{tab:Askey scheme}
\end{table}

For the stochastic process $\mathbf{x}=\mathbf{x}(t;Z)$, subject to the dynamics in \eqref{eq:general nonlinear uncertain system}, the gPC expansion reads
\begin{equation}\label{eq:gPC expansion random process}
\begin{array}{lll}
    &\mathbf{x}(t;Z) = \sum_{\alpha \in \mathbb{N}_0^d} \hat{\mathbf{x}}_\mathbf{\alpha}(t)\Phi_\mathbf{\alpha}(Z), \\
    &\hat{\mathbf{x}}_\mathbf{\alpha}(t) =\begin{bmatrix}
    \hat{\mathbf{x}}_{\alpha,1}(t)& \dots & \hat{\mathbf{x}}_{\alpha,n}(t)
    \end{bmatrix}^{\top}\in \mathbb{R}^n,\\
    &\hat{\mathbf{x}}_{\alpha,j}(t)\coloneq\mathbb{E}[\mathbf{x}_j(t,\cdot)\Phi_\mathbf{\alpha}], \,\, j=1,\dots, n.
\end{array}
\end{equation}

Intuitively, gPC methods represent the stochastic process $\mathbf{x}(t;Z)$ as a series expansion with respect to an appropriate basis of orthogonal polynomials depending on the uncertain parameters $Z$ and the corresponding density $\rho_Z$, with coefficients depending on time $t$, as in equation \eqref{eq:gPC expansion random process}. 
The spectral coefficients $\left\{\hat{\mathbf{x}}_\mathbf{\alpha}(t) \right\}_{\alpha}$ contain all the temporal information, while the stochasticity is confined in the orthogonal polynomials $\left\{ \Phi_\mathbf{\alpha}(Z)\right\}_{\alpha}$, thereby achieving separation of the stochastic and deterministic elements that define the dynamics. The relation \eqref{eq:orthogonality of polynomials} and the linearity of the spectral representation \eqref{eq:gPC expansion random process} can thus be used to compute the summary statistics of $\mathbf{x}(t;Z)$. For arbitrary statistical moments, we refer the reader to \cite{lefebvre2020moment}. In this work we only consider mean and variance, extracted as
\begin{equation}
\mu_\mathbf{x}(t) = \hat{\mathbf{x}}_\mathbf{0}(t) \quad \mbox{and} \quad \sigma_\mathbf{x}^2(t) = \sum_{|\mathbf{\alpha}| \neq 0} \hat{\mathbf{x}}_\mathbf{\alpha}^2(t).
\label{eq:mean and variance with gPC}
\end{equation}

The expansion \eqref{eq:gPC expansion random process} holds theoretically; in practice, for numerical implementation, it must be truncated to obtain a finite-sum approximation that can be substituted into the expression of the statistical moments in equations \eqref{eq:mean and variance with gPC}.  Finite-sum approximations can be computed either by intrusive approaches (e.g. Galerkin), which modify the governing equations of the original system by truncating the gPC expansion, or by non-intrusive approaches (e.g. collocation) that treat the model as a black box and use sampling to estimate the spectral coefficients indirectly, e.g. via least-squares regression. 

As a first approach to generate an approximation of $\mathbf{x}(t;Z)$, assume (only for demonstration purposes, without loss of 
generality) that \eqref{eq:general nonlinear uncertain system} is a polynomial system, i.e. 
\begin{equation}\label{eq:fPoly}
\begin{array}{lll}
&f(\mathbf{x},Z) = \sum_{|k|\leq L}a_k(Z)x_1^{k_1}(t,Z)\ldots x_n^{k_n}(t,Z),\\
&a_k(Z) = \sum_{\alpha \in \mathbb{N}_0^d}\hat{a}_{k,\alpha} \Phi_{\alpha}(Z)\in \mathbb{R}^n,
\end{array}
\end{equation}
where $k=(k_1,\dots,k_n)\in \mathbb{N}_0^n$ and $L\in \mathbb{N}_0$. By substituting \eqref{eq:gPC expansion random process} and \eqref{eq:fPoly} into \eqref{eq:general nonlinear uncertain system}, multiplying both sides by $\Phi_{\beta}$ and taking the expectation, we have
\begin{equation}\label{eq:AccurateODEsProj}
\begin{array}{lll}
&\frac{d}{dt} \hat{\mathbf{x}}_{\beta}(t) = \mathbb{E}[f(\sum_{\alpha \in \mathbb{N}_0^d} \hat{\mathbf{x}}_\mathbf{\alpha}(t)\Phi_\mathbf{\alpha},Z)\Phi_{\beta}]\\
&\hspace{12mm}=\sum_{|k|\leq L}\mathbb{E} \left[\chi_k \Phi_{\beta}\right],\ \beta \in \mathbb{N}_0^d,\\
&\chi_k \coloneq \left(\sum_{\alpha \in \mathbb{N}_0^d}\hat{a}_{k,\alpha} \Phi_{\alpha} \right)\prod_{j=1}^n \left(\sum_{\alpha\in \mathbb{N}_0^d} \hat{\mathbf{x}}_{\alpha,j}(t)\Phi_{\alpha}\right)^{k_j}.
\end{array}
\end{equation}

The approximation to $\mathbf{x}(t;Z)$ is then of the form 
\begin{equation}\label{eq:gPC expansion random process 2}
\begin{array}{lll}
    &\mathbf{x}_N(t;Z) = \sum_{|\alpha|\leq N}\tilde{\mathbf{x}}_{\alpha}(t)\Phi_{\alpha}(Z), \\
    &\tilde{\mathbf{x}}_{\alpha}(t) = \begin{bmatrix}
    \tilde{\mathbf{x}}_{\alpha,1}(t)& \dots & \tilde{\mathbf{x}}_{\alpha,n}(t)
    \end{bmatrix}^{\top} \in \mathbb{R}^n, 
\end{array}
\end{equation}
where the coefficients $\left\{\tilde{\mathbf{x}}_i(t) \right\}_{|\alpha| \leq N}$ in \eqref{eq:gPC expansion random process 2} are obtained as solutions to the truncated system: 
\begin{equation}\label{eq:ApproxODEsProj}
\begin{array}{lll}
&\frac{d}{dt} \tilde{\mathbf{x}}_{\beta}(t) = \sum_{|k|\leq L}\mathbb{E} \left[\tilde{\chi}_k \Phi_{\beta}\right],\ |\beta| \leq N, \\
&\tilde{\chi}_k \coloneq \left(\sum_{|\alpha| \leq N}\hat{a}_{k,\alpha} \Phi_{\alpha} \right)\prod_{j=1}^n \left(\sum_{|\alpha| \leq N} \tilde{\mathbf{x}}_{\alpha,j}(t)\Phi_{\alpha}\right)^{k_j}.
\end{array}
\end{equation}
System \eqref{eq:ApproxODEsProj} is obtained from system \eqref{eq:AccurateODEsProj} by retaining only polynomials whose degree is less or equal to $N$. This approach is the so-called stochastic Galerkin projection.

A competing approach to generate surrogates of $\mathbf{x}(t;Z)$ is the collocation method, based on sampling the values of the random variable $Z$. Let $\left\{Z^{(j)}\right\}_{j=1}^{S_C}\subset \mathcal{I}_{Z},\ S_\text{C}\geq \binom{N+d}{d}$, be a grid of nodes sampled in the range of $Z$. There are multiple ways to generate such a grid: as pseudo-random Monte Carlo sampling according to $\rho_Z$, or as nodes of a pre-selected quadrature rule, e.g. the Smolyak sparse grid \cite{smolyak1963quadrature} or a Clenshaw-Curtis grid \cite{engels1980numerical}. For each $1\leq j\leq S_\text{C}$, let $\bar{\mathbf{x}}^{(j)}(t)$ be a solution to the \emph{deterministic} system \eqref{eq:general nonlinear uncertain system} with $Z$ replaced by the value $Z^{(j)}$. The desired approximation to $\mathbf{x}(t;Z)$ is considered again in the form \eqref{eq:gPC expansion random process 2}, subject to the interpolation conditions
\begin{equation}\label{eq:InterpCond}
\begin{array}{lll}
\mathbf{x}_N(t;Z^{(j)}) = \sum_{|\alpha|\leq N}\tilde{\mathbf{x}}_{\alpha}(t)\Phi_{\alpha}(Z^{(j)})=\bar{\mathbf{x}}^{(j)}(t)
\end{array}
\end{equation}
where $1\leq j \leq S_\text{C}$. These conditions impose an over-determined (due to the condition on $S_\text{C}$) linear system on the coefficients $\left\{\tilde{\mathbf{x}}_{\alpha}(t)\right\}_{|\alpha|\leq N}$, which can be solved either explicitly or via a least-squares projection.

\paragraph{Remark}
The collocation procedure has the same form regardless of the structure of the dynamics. Conversely, the Galerkin approach can be made more precise for polynomial systems when expectations of higher order products of basis elements are explicitly available  \cite{Petzke2020_PoCET}, and can be embedded in \eqref{eq:ApproxODEsProj}, without any additional approximation (such as cubature formulas for integral approximations). 
$\hfill\diamond$ \\

The rate of convergence of the approximation of a static map $f(Z) = \sum_{\alpha \in \mathbb{N}_0^{d}} f_{\alpha}\Phi_{\alpha}(Z)$ depends on the smoothness of $f$ and the type of orthogonal polynomial basis functions $\{\Phi_\alpha \}_\alpha$ employed in the approximation \cite{xiu2010numerical}. Let $H^{\ell}_{\rho_Z}(\mathcal{I}_Z)\subseteq L^2_{\rho_Z}(\mathcal{I}_Z)$, $\ell\in \mathbb{N}_0$, denote the Sobolev space of functions $g:\mathcal{I}_Z\to \mathbb{R}$ having (weak) derivatives $D^{\beta}g\in L^2_{\rho_Z}(\mathcal{I}_Z)$, where $\beta = (\beta_1,\dots,\beta_d)\in \mathbb{N}_0^d$, $\left|\beta \right| = \beta_1+\dots+\beta_d \leq \ell$  and $D^{\beta} = \partial_ {z_1}^{\beta_1}\dots \partial_ {z_d}^{\beta_d}$ is a differentiation operator. If $f\in H^{\ell}_{\rho_Z}(\mathcal{I}_{Z})$ with $\ell\in \mathbb{N}$,
the approximation error $\lVert f - \sum_{|\alpha|\leq N}f_{\alpha}\Phi_{\alpha}\rVert_{L^2_{\rho_Z(\mathcal{I}_Z)}}$ is $\mathcal{O}(N^{-\ell})$. For the case of $\ell=0$, i.e. $H^{\ell}_{\rho_Z}(\mathcal{I}_{Z})=L^2_{\rho_Z}(\mathcal{I}_Z)$, convergence is guaranteed without a prescribed convergence rate.
For analytic $f$, the convergence rate is exponential, i.e. $\mathcal{O}(e^{-\sigma N})$ for some constant $\sigma > 0$. For solutions to ODE systems \eqref{eq:general nonlinear uncertain system}, few rigorous results that quantify the convergence rates exist. Some analytic results and numerous numerical studies in the literature indicate that similar convergence rates hold for $\left\|\mathbf{x}(t;\cdot)-\mathbf{x}_{N}(t;\cdot) \right\|_{L^2_{\rho_z}(\mathcal{I}_Z)}$, at least on compact time intervals; see \cite{kim2013generalised,xiu2002wiener} for more details. 

As analytical frameworks are missing to estimate the convergence of gPC methods on general systems and to compare them with Monte Carlo convergence results, numerical investigation should be employed to gain insight as to how gPC compares with MC methods. Such numerical comparisons have been explored rarely and only few case-specific results are available, especially for systems in biology \cite{SonDu2020}, where Monte Carlo has been the main workhorse for decades. To perform a detailed comparison of gPC variants and Monte Carlo methods in neuroscience models, characterised by different operating regimes emerging from their dynamics, we consider as case studies the Hindmarsh-Rose model for single-neuron dynamics, the Jansen-Rit model for collective neural dynamics, and the Epileptor model for epilepsy. 



\subsection{The Hindmarsh-Rose (HR) model}

The Hindmarsh-Rose (HR) model \cite{hindmarsh1984model} captures the dynamics of action-potential within a single neuron capable of bursting activity \cite{innocenti2007dynamical}. It can display numerous patterns of trajectory behaviours (regimes), ranging from cyclic spiking to bursting or chaos, depending on the considered parameter sets \cite{montanari2022functional}, which are often uncertain due to experimental design or neural activity. The corresponding non-dimensionalised set of ODEs is
\begin{equation}
    \begin{cases}
    \dot{x} = y - a x^3 + bx^2 - z + I, \\
    \dot{y} = c - dx^2 - y, \\
    \dot{z} = r[s(x - x_R) - z],
    \end{cases}
    \label{eq:HR}
\end{equation}
where $x$ represents membrane potential, $y$ is a fast recovery current, $z$ is a slow adaptation current; $I$ is an externally applied current (either experimental current injection or in-vivo synaptic current), and the other terms are model parameters. Their values can be inferred from fitting the model to experimental data \cite{gu2013biological}, and may depend on the considered organism and come with significant uncertainties. If parameter values are changed \emph{deterministically}, bifurcation studies \cite{gonzalez2007complex, storace2008hindmarsh} allow to identify parameter combinations corresponding to various regimes.



\begin{figure}[ht!]
	\centering
	\includegraphics[width=\linewidth]{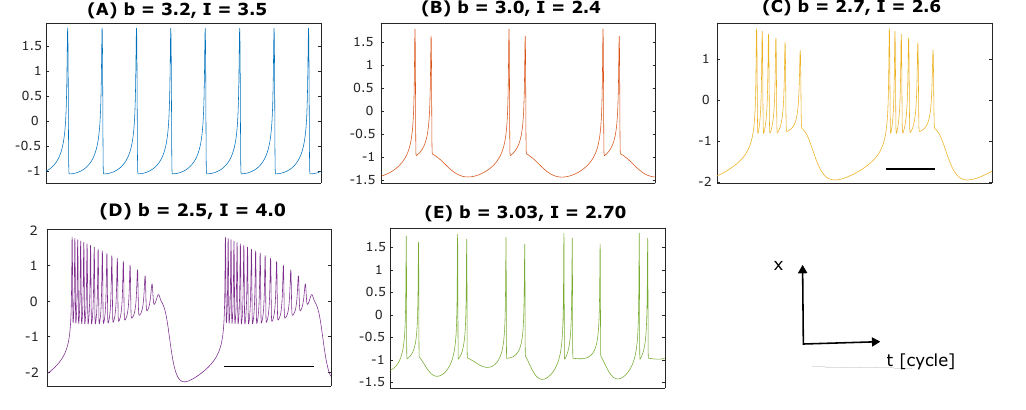}
	\caption{\small Examples of time series $x(t)$ for the regimes ($A$)--($E$) of the HR model \eqref{eq:HR}, described in the main text. The lines in ($C$) and ($D$) identify a burst of spikes.}
	\label{fig:examples}
\end{figure}

Fixing the parameters $a$, $c$, $d$, $r$, $s$ and $x_R$ in \eqref{eq:HR} to default values $a = 1$, $c = 1$, $d = 5$, $r = 0.01$, $s = 4$, $x_R = -8/5$,  and letting $I \in [2.2, 4.4]$ and $b \in [2.5, 3.3]$ vary within the given intervals, the HR dynamics can give rise to five different regimes \cite{de2008predicting, gonzalez2007complex, storace2008hindmarsh}, as illustrated in Fig.~\ref{fig:examples}: ($A$) quiescence, ($B$) tonic spiking, ($C$) square-wave bursting, ($D$) plateau bursting and ($E$) chaotic bursting. 

\subsection{The Jansen-Rit (JR) model}

The Jansen-Rit model \cite{jansen1995electroencephalogram} is a neural mass model explaining the dynamics of interconnected neuronal populations of different neural types (pyramidal cells and inhibitory and excitatory inter-neurons in the cortical area). Similarly to HR, it can exhibit regimes such as quiescence, chaos, periodic and quasi-periodic behaviours, for biologically plausible parameter ranges. The model is given by
\begin{equation}
    \begin{cases}
        \dot{y}_1 = y_4 , \quad  \dot{y}_2 = y_5 , \quad \dot{y}_3 = y_6 , \\
        \dot{y}_4 = A a S(y_2 - y_3) - 2a y_4 - a^2 y_1 , \\
        \dot{y}_5 = A a [p + C_2 S(C_1 y_1)] - 2a y_5 - a^2 y_2 , \\
        \dot{y}_6 = B b C_4  S(C_3 y_1) - 2 b y_6 - b^2 y_3 ,
    \end{cases}
    \label{eq:JR}
\end{equation}
where $S(V) = \nu_{M} / \left(1+e^{r(V_0 - V)}\right)$, with $\nu_{M}$ being the maximal firing rate of the family of neurons and $V_0$ the excitability threshold of the population; $r$ is the \textit{firing threshold variance}, i.e., the slope of the sigmoid at $V_0$. All connectivity constants $C_i$ are proportional to a single parameter $C$: $C_1 = C$, $C_2 = 0.8  \ C$, $C_3 = 0.25 \ C$, $C_4 = 0.25 \ C$ \cite{jansen1995electroencephalogram}. Variables $y_i$ represent neural potentials, and the model parameters are usually fixed to reproduce basic properties of postsynaptic potentials and cortical activity. Typical values, which we employ for simulations unless it is specified otherwise, are $C =  135$, $A= 3.25$ mV, $B= 22$ mV, $a=  100$ s$^{-1}$, $b=  50$ s$^{-1}$, $V_0 = 6$ mV, $\nu_{max} = 5$ s$^{-1}$, $r= 0.56$ mV$^{-1}$; $p$ is usually uncertain within a maximum interval $[0,400]$ s$^{-1}$ \cite{jansen1995electroencephalogram}.

\subsection{The Epileptor model}

The Epileptor model \cite{jirsa2014nature} is a phenomenological mean-field model aimed at capturing the average seizure dynamics of blocks of neurons. It expands the HR model \eqref{eq:HR} by adding couplings to generate seizure-like events. The non-dimensionalised model is given by
\begin{equation}
\begin{cases}
\dot x_1 = x_2 - f_1(x_1,x_4) - x_3 + I_{1} ,\\
\dot x_2 = r_2 - 5x_1^2 - x_2 ,\\
\dot x_3 = \frac{1}{\tau_0} (4(x_1 - r_1) - x_3) ,\\
\dot x_4 = -x_5 + x_4 - x_4^3 + I_{2} + 2u - 0.3(x_3 - 3.5) ,\\
\dot x_5 = \frac{1}{\tau_2} (-x_5 + f_2(x_4)),\\
\dot{u} = -\gamma (u-0.1 x_1) ,
\end{cases}
\label{eq:Epileptor}
\end{equation}
where $(r_{1}, r_{2}, I_{1}, I_{2}) = (-1.6, 1, 3.1, 0.42)$ are the system parameters, $(\tau_0,\tau_2,\gamma) = (2857, 10, 0.01)$ are timescale constants, $u$ is a low-pass filter ``dummy variable'' \cite{jirsa2014nature} and the non-smooth coupling functions are given by
\begin{equation}
    \begin{aligned}
        f_1(x_1,x_4) &= 
        \begin{cases}
        x_1^3 - 3x_1^2, \quad & x_1 < 0, \\
        (x_4 - 0.6(x_3 - 4)^2)x_1, \quad & x_1 \geq 0, \\
        \end{cases} \\
        f_2(x_4) &= 
        \begin{cases}
        0, \quad & x_4 < -0.25, \\
        6(x_4 + 0.25), \quad & x_4 \geq -0.25. \\
        \end{cases}
    \end{aligned}
\end{equation}
Variables $x_1$ and $x_2$ govern the oscillatory behaviours, while $x_4$ and $x_5$ introduce the spikes and the wave components typical of seizure-like events, and $x_3$ is a slow permittivity variable driving the system to the seizure threshold. 

\section{Results: gPC for Neurological Models}

\subsection{Methodology}
We aim to assess the performance of gPC methods in reconstructing summary statistics for individual regimes of each model. To this end, we select parameters of interest $\hat{p}_i$, determined from a literature overview and based on the biological meaning of each model. For the $i$th parameter in the $j$th regime, we consider a uniform distribution, i.e., $\hat{p}_i \sim \mathcal{U}([\hat{p}^\text{min}_{i, j}, \hat{p}^\text{max}_{i, j}])$. Minimum and maximum values may differ among regimes, depending on their persistence set; see e.g. \cite[Figure 3]{barrio2011parameter}. For instance, in the HR model, for regimes $j \in \{A, B, C\}$ we fix $I=I_{\text{HR}, j}$ and vary $b \sim \mathcal{U}([b^\text{min}_{\text{HR}, j}, b^\text{max}_{\text{HR}, j}])$; for regime $D$, we fix $b=b_{\text{HR}, D}$ and vary $I \sim \mathcal{U}([I^\text{min}_{\text{HR}, D}, I^\text{max}_{\text{HR}, D}])$. Chaos (regime $E$) is currently not investigated. Choices of parameter values and intervals for the various models are reported in the figure captions; the \textsc{Matlab} code used to generate all our results is available at {\small{\texttt{https://github.com/Uros-S/UQgPC-neuroscience}}}

The computational complexity is assessed through the use of a proxy measure, the running time $\tau_m$, where $m$ refers to the considered surrogate method (Monte Carlo, Galerkin or collocation), on a fixed hardware (a Dell Inspiron 16 laptop with 16 GB RAM and 1.90 GHz Intel i5-1340P core, running Windows 11 and kept on charge, to avoid power saving modes). To assess the accuracy performance of gPC methods, we first construct a ``benchmark'' for each model and regime, employing a Monte Carlo (MC) scheme with $N_{\text{MC}} = 10^5$ samples that guarantees accuracy $\varepsilon < 0.01$ for a confidence $\delta = 0.01$ \cite{tempo2013randomized} albeit requiring $\tau_{\text{MC}} > 7$h for the HR model and even $\tau_{\text{MC}} > 29$h for the Epileptor model. 

Given a final simulation time $T$ and a time-step $dt$ such that $T = N_e \ dt$, with $N_e \in \mathbb{N}$, we compute the error vector $e \in \mathbb{R}^{N_e}$ between the MC-based ``benchmark'' and the results obtained with the surrogate models. Deviations for both mean and variance are estimated using the element-wise root mean square error (RMSE)
\begin{equation*}
    e_{\text{RMS}} = \sqrt{\frac{1}{N_e}\sum_{n=1}^{N_e} e_n^2},
\end{equation*}
where $e_n$ is the $n$th entry of vector $e$.

As a representative MC approximated method, we use a Monte Carlo chain with $N_{\text{MC}} = 5000$ samples; in our analysis, we always consider its best (either the fastest or the most precise) run, among those from all considered regimes.


For the Galerkin approach, we assess the accuracy of the obtained surrogate model \eqref{eq:gPC expansion random process 2} with respect to the expansion order $N=N_\text{G}$. For the collocation approach, we assess the accuracy of the obtained surrogate model \eqref{eq:gPC expansion random process 2} with respect to both the expansion order $N=N_\text{C}$ and the number of collocation samples $S_\text{C}$. 
The time $\tau_m$ is estimated from the first call of the polynomial projection to the estimation of the statistical moment of interest. This is a worst-case time in the case of new simulations, as it includes the compilation time for the expansion polynomials, which can then be stored, without concurring to additional processing in the case of repeated runs.  

Simulations of all models are performed using \textsc{Matlab} \texttt{ode45} function. Consistently with studies in literature, we use $dt = 0.01$ and $T=1200$ for the HR model, $dt = 2.5\cdot10^{-4}$ and $T=2.5$ for the JR model, and $dt = 0.01$ and $T=4500$ for the Epileptor model; initial conditions are set at zero for the HR and JR models, and at $[0 \ -5 \ 3 \ 0 \ 0 \ 0]^\top$ for the Epileptor model. 
gPC approaches are implemented via the PoCET \textsc{Matlab} toolbox \cite{Petzke2020_PoCET}, already optimised for uncertainty quantification. To maintain consistency and fairness of comparison, Monte Carlo simulations are also performed using PoCET built-in functions.

For the JR model and the Epileptor model, which are not polynomial, the Galerkin method is not applicable with the PoCET toolbox; therefore, to maintain consistency, for these two models we only employ the collocation method.

\subsection{gPC methods outrun MC and maintain accuracy}

\textbf{HR model.} The rapid and approximated MC chain with $N_{\text{MC}}=5000$ has running time in the order of $10^3$ s, and mean and variance of RMSE are around $0.01$.

\begin{figure*}[ht!]
	\centering
	\includegraphics[width=\linewidth]{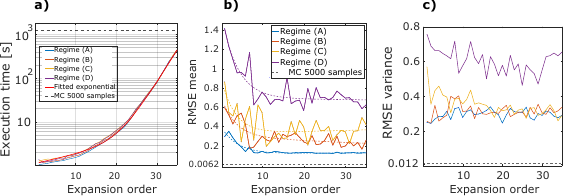}
    \vspace{-3mm}
	\caption{\small Galerkin gPC approximation, for the considered regimes of the HR model \eqref{eq:HR}, depending on the expansion order $N_\text{G}$. For regime (A), $I=3.5$ and $b\sim\mathcal{U}([3.1,3.3])$; (B), $I=2.4$ and $b\sim\mathcal{U}([3.0,3.15])$;  (C), $I=2.6$ and $b\sim\mathcal{U}([2.6,2.8])$; (D), $b=2.5$ and $I\sim\mathcal{U}([3.8,4.2])$. Gray dashed lines: values corresponding to the best run of the MC approximation (either the fastest or the most precise, among those for all considered regimes), with $\tau_{\text{MC}} = 1238$ s, mean of the RMSE $= 0.0062$, variance of the RMSE $=0.012$. (a) Running time $\tau_\text{G}$. Fitted exponential \eqref{eq:galerkin_fit} for regime $B$. (b) Mean of the RMSE for the variable $x$. Dashed lines represent empirical exponential fitting from \eqref{eq:galerkin_fit}. (c) Variance of the RMSE for  $x$.}
	\label{fig:galerkin}
\end{figure*}

For the Galerkin approximation, the computation time $\tau_{\text{G}}$ in all regimes scales sub-exponentially for small expansion orders and exponentially for medium-large $N_\text{G}$, following a best fit relation 
\begin{equation}
    \tau_\text{G} = \tilde{a} e^{\tilde{b} N_\text{G}} + \tilde{c} e^{\tilde{d} N_\text{G}} , 
    \label{eq:galerkin_fit}
\end{equation}
with $\tilde{a},\tilde{b},\tilde{c},\tilde{d} \in \mathbb{R}$ (see also Fig.~\ref{fig:galerkin}a). For any considered expansion order, $\tau_\text{G}$ is at least one order of magnitude lower than $\tau_{\text{MC}}$, marking a significant acceleration in computation. Despite fluctuations related to stochasticity, the mean RMSE decays rapidly (following the same trend as in \eqref{eq:galerkin_fit}, with different parameters) as the expansion order increases (Fig.~\ref{fig:galerkin}b). For regimes $A$ and $B$, the mean error with high $N_\text{G}$ is close to that of the MC chain, while the mean error is significantly higher for regimes with bursting ($C$ and $D$). Overall, the variance of the RMSE remains approximately constant for increasing $N_\text{G}$, mostly around values of $0.3$, which is about $30$ times that of MC; see Fig.~\ref{fig:galerkin}c. These results suggest that the Galerkin approximation can drastically speed up the computation time with respect to MC methods, but at the price of decreasing accuracy, which might be acceptable in certain applications, such as the preliminary sweeping of the parameter space in order to identify regions of interest.

\begin{figure*}[ht!]
	\centering
	\includegraphics[width=\linewidth]{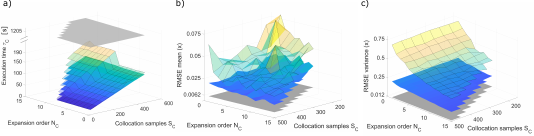}
    \vspace{-8mm}
	\caption{\small Collocation gPC approximation for the HR model \eqref{eq:HR}. We show the best-case and worst-case surfaces, which interpolate minimum and maximum values, for each point $\{N_\text{C},S_\text{C}\}$, of $\tau_\text{C}$ and of performance metrics among the four cases (one per regime) considered in Fig.~\ref{fig:galerkin}. Gray surfaces: best-case values of the MC approximation, with $\tau_{\text{MC}} = 1238$ s, mean of the RMSE $= 0.0062$, variance of the RMSE $=0.012$. (a) Running time $\tau_\text{C}$; note that the $z$-axis is cut to show the MC surface, with very high $\tau_\text{C}$. (b) Mean of the RMSE for the variable $x$. (c) Variance of the RMSE for the variable $x$.}
	\label{fig:collocation}
\end{figure*}

Collocation requires to set both the expansion order $N_\text{C}$ and the number of collocation samples $S_\text{C}$. Fig.~\ref{fig:collocation} shows the best-case and worst-case surfaces obtained from testing all regimes: given for each $\{N_\text{C},S_\text{C}\}$ a vector $z(N_\text{C},S_\text{C}) = [z_i(N_\text{C}, S_\text{C})]$, where $i$ denotes the regime, we show the surfaces that interpolate $\max(z)$ and $\min(z)$. For comparison, we also show the surface associated with MC values, in gray. The running time $\tau_\text{C}$ (Fig.~\ref{fig:collocation}a) and the mean RMSE (Fig.~\ref{fig:collocation}b) are more significantly influenced by the number of collocation samples, whereas the expansion order has relatively little impact, at least in the considered interval. On the other hand, a higher expansion order significantly improves the variance of the RMSE (Fig.~\ref{fig:collocation}c), which is consistent with \eqref{eq:mean and variance with gPC}. Notably, even for high $N_\text{C}$ and high $S_\text{C}$, the running time remains around 2 minutes, while MC takes about 20 minutes, highlighting a $10\times$ acceleration. We do not consider expansion orders above $N_\text{C} > 15$, which would further slow down the computation, since the mean of the RMSE is already sufficiently close to that obtained with MC (only twice its value) even with a relatively low expansion order $N_\text{C}$; the variance of the RMSE is however one order of magnitude larger than that obtained with MC. \\

\begin{figure*}[ht!]
	\centering
	\includegraphics[width=\linewidth]{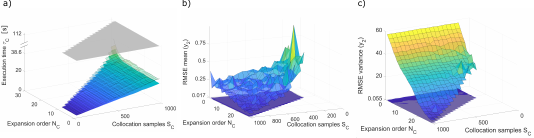}
    \vspace{-8mm}
	\caption{\small Collocation gPC approximation for the JR model \eqref{eq:JR}. We show the best-case and worst-case surfaces, which interpolate minimum and maximum values, for each point $\{N_\text{C},S_\text{C}\}$, of $\tau_\text{C}$ and of performance metrics among six different cases with $C \in \{68, 95, 128, 135, 155, 173\}$, and $p \sim \mathcal{U}([120,320])$. Gray surfaces: best-case values of the MC approximation, with $\tau_{\text{MC}} = 112$ s, mean of the RMSE $= 0.017$, variance of the RMSE $=0.055$. (a) Running time $\tau_\text{C}$; note that the $z$-axis is cut to show the MC surface, with very high $\tau_\text{C}$. (b) Mean of the RMSE for the variable $y_2$. (c) Variance of the RMSE for the variable $y_2$.}
	\label{fig:collocationJR}
    \vspace{-4mm}
\end{figure*}

\textbf{JR model.} We assume that the uncertain parameter $p$ is uniformly distributed, $p \sim \mathcal{U}([120,320])$, and we test different dynamical cases by setting six different values of $C \in \{68, 95, 128, 135, 155, 173\}$, so that the dynamical regimes resemble cortical signals that are non-oscillatory, waning sinusoidal-like, sinusoidal-like and low frequency quasi-periodic; see \cite[Figure 3]{jansen1995electroencephalogram}. 
For simulations with MC, $\tau_{\text{MC}}$ is in the order of $10^2$ s. Collocation achieves even better performance, providing statistics with a running time below $40$ s. Similarly to the HR case, Fig.~\ref{fig:collocationJR} shows the best- and worst-case surfaces for $\tau_\text{C}$ and the RMSE statistics. The results are consistent with the HR case study: collocation speeds up the computation, and a higher number of collocation samples $S_\text{C}$ significantly improves the RMSE mean, while the RMSE variance is mostly affected by the expansion order $N_\text{C}$, and decreases rapidly (for the worst-case surface) as $N_\text{C}$ increases. Notably, the best collocation approximations are \emph{\textbf{even more accurate}} than the best ones obtained with MC; see the blue surface overlapping with the gray one in Fig.~\ref{fig:collocationJR}b,c. Hence, in some cases,
the collocation method provides precise approximations for most $\{N_\text{C}, S_\text{C} \}$ combinations. The discrepancy between best-cases and worst-cases is drastically reduced by selecting $N_\text{C} > 15$ and $S_\text{C} > 500$.\\

\begin{figure*}[t!]
	\centering
	\includegraphics[width=\linewidth]{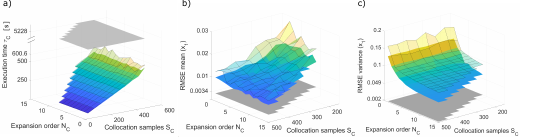}
    \vspace{-8mm}
	\caption{\small Collocation gPC approximation for the Epileptor model \eqref{eq:Epileptor}. We show the best-case and worst-case surfaces, which interpolate minimum and maximum values, for each point $\{N_\text{C},S_\text{C}\}$, of $\tau_\text{C}$ and of performance metrics among four different cases: $I_1 \sim \mathcal{U}([3.1,3.6])$ with $I_2=0.42$; $I_1 \sim \mathcal{U}([3.1,3.6])$ with $I_2=0.2$; $I_2 \sim \mathcal{U}([0.2,0.35])$; $\tau_0 \sim \mathcal{U}([2500,2750])$. Gray surfaces: best-case values of the MC approximation, with $\tau_{\text{MC}} = 5228$ s, mean of the RMSE $= 0.0062$, variance of the RMSE $=0.012$. (a) Running time $\tau_\text{C}$; note that the $z$-axis is cut to show the MC surface, with very high $\tau_\text{C}$. (b) Mean of the RMSE for the variable $x_1$. (c) Variance of the RMSE for the variable $x_1$.}
	\label{fig:collocationEpi}
    \vspace{-2mm}
\end{figure*}

\textbf{Epileptor model.} We consider
the  single dynamical regime that reproduces seizure-like dynamics, so as to be in line with the model's goal and meaning. Bounds of the uniform distribution for $\hat{p}_i$ are thus chosen not to yield other regimes. To construct the vector $z(N_\text{C},S_\text{C})$, three different realisations are obtained by considering, one at a time, uniform distributions for $I_1 \sim \mathcal{U}([3.1,3.6])$, $I_2 \sim \mathcal{U}([0.2,0.35])$, $\tau_0 \sim \mathcal{U}([2500,2750])$, while leaving all the other parameters at their nominal value; a fourth realisation considers $I_2=0.2$ and $I_1 \sim \mathcal{U}([3.1,3.6])$. Since the Epileptor requires long simulations with small time steps, to capture fast discharges, spikes and wave events \cite{jirsa2014nature}, the computing time for the MC exceeds $5000$ s, while the gPC collocation method requires one order of magnitude less and $\tau_\text{C}$ scales almost linearly with $S_\text{C}$; see Fig.~\ref{fig:collocationEpi}a. Results for the RMSE mean and variance are similar to those of the HR, as the RMSE mean becomes very close to that obtained with MC (twice its value) even with a relatively low expansion order $N_\text{C}$, while the RMSE variance is one order of magnitude larger. The analogy with the results achieved with the HR model is not surprising, as the Epileptor uses the HR model as its kernel; the main difference lies in the small discrepancy between best- and worst-case surfaces for the RMSE variance, which is expected since we considered the very specific regime of seizure-like events.

\subsection{Computing probabilistic robustness}

Our findings support the use of collocation gPC methods to study probabilistic robustness, achieving a good compromise between computational time and accuracy of the results. With respect to Galerkin, collocation methods achieve higher accuracy in all regimes. On the other hand, MC is too computationally demanding when investigating multiple uncertainties or considering large parametric spaces (which are common in neuroscience), whereas we have shown that gPC is not. To analyse regime robustness for the HR model, we therefore employ collocation gPC. A high $N_\text{C}$ yields better RMSE performance without an excessive increase in $\tau_\text{C}$. For $S_\text{C}$, we notice in Fig.~\ref{fig:collocation} that $S_\text{C} > 400$ does not improve RMSE statistics, but only increases $\tau_\text{C}$. Hence, we select $S_\text{C} =$ 400 and $N_\text{C} = 15$.

As a first case study, to showcase the possibility of using gPC for probabilistic robustness, we consider a selected regime: plateau bursting ($D$) of the HR. For a given $I=4.2$, we thus investigate the probability that $D$ persists, depending on sampled values $b \sim \mathcal{U}(\mathcal{I}_b)$, with $\mathcal{I}_b = [2.4, 2.8]$, where $2.4$ is associated with typical leftmost values considered in phase planes from the literature \cite{barrio2011parameter}, and $\Delta b = 0.4$ is informed by experimental uncertainties obtained from fitting experimental data \cite{de2008predicting}. 
Plateau bursting is known to be characterised by a Hopf bifurcation \cite{gonzalez2007complex} marking the end of the burst of spikes at $x = 0$. We use this signature on the gPC mean to automatically detect whether the simulations maintain regime $D$ or not. We further characterise the proportion $P^* = \frac{\mbox{size}(A_b)}{\mbox{size}(\mathcal{I}_b)}$, where $A_b \subseteq \mathcal{I}_b$ is the interval for which $D$ is preserved, as a \textbf{\textit{probabilistic measure of robustness}}.

\begin{figure}[hb]
	\centering
	\includegraphics[width=0.5\linewidth]{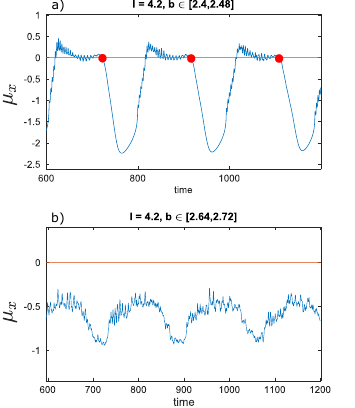}
	\caption{\small Mean $\mathbb{E}[x] = \mu_x(t)$, computed using the collocation method with $N_\text{C} = 15$ and $S_\text{C} = 400$, for system \eqref{eq:HR} with $I=4.2$ and $b$ uniformly distributed in given intervals: (a) $b \sim \mathcal{U}([2.4, 2.48])$, fully within regime $D$, and (b) $b \sim \mathcal{U}([2.64, 2.72])$, fully outside regime $D$. For regime $D$, the red circles mark the end of the bursts of spikes at $\mu_x = 0$.}
	\label{fig:prob_rob}
\end{figure}

The gPC mean in each interval $A_b \subseteq \mathcal{I}_b$ is immediately computed in PoCET following equation \eqref{eq:mean and variance with gPC}. Fig.~\ref{fig:prob_rob}a shows an example of the mean $\mu_x(t)$ for
$b \sim \mathcal{U}([2.4, 2.48])$, for which the burst of spikes ends at $\mu_x(t) = 0$ (red dot, when the neuron re-polarises). Fig.~\ref{fig:prob_rob}b shows an example for $b \sim \mathcal{U}([2.64, 2.72])$, whose re-polarisation occurring at $\mu_x(t) < 0$ clearly indicates a signature of regime $C$ (\textit{cf.} Fig.~\ref{fig:examples}), and not $D$. Other intervals $A_b$ are systematically checked using the same criterion, to study the persistence of $D$. 

We eventually identify in $A_b^* = [2.4, 2.56]$ the largest robust interval for regime $D$, corresponding to $P^* = 40$\%, which is consistent with biological studies \cite{de2008predicting} suggesting that, in experiments, square-wave bursts are more common than plateau bursting. The result for $A_b^*$ is in line with studies in the literature using alternative methods \cite{storace2008hindmarsh, barrio2011parameter, gonzalez2007complex}. The whole process took about 8.2 minutes to complete, with clear advantages over MC in terms of computational time (recall that a \textit{single} MC run on an arbitrary distribution of $b$ takes around 20 minutes). Our proposed method thus allows for a rapid analysis that is sufficient for the probabilistic quantification of persistence of a desired regime. Further refinements may be used to hone the probability estimates.

\section{Concluding Discussion}

We have considered gPC methods as suitable candidates to accelerate parametric uncertainty and probabilistic robustness studies in three important models in neuroscience, achieving a significant increase in computational efficiency with respect to MC approaches and, in some cases, comparable accuracy.

Both intrusive and non-intrusive gPC approaches have been considered, and collocation has been identified as an overall better approach for the study of neural dynamics.

In particular, for complex systems in neuroscience, which are characterised by multiple distinct regimes, the collocation method offers a good accuracy and a significant reduction in running time compared to MC. 
Conversely, the Galerkin approximation, which is the most computationally efficient, yields a drop in accuracy that may hinder the investigation of neural dynamics. 
So, we have focused on the collocation approach. By identifying how the statistics of the RMSE and of the computational time depend on the polynomial order and the number of collocation points, our results also demonstrate the tradeoff between efficiency and accuracy, depending on the desired study objective.

Our numerical findings reveal that, in some cases, a higher computational efficiency of the gPC approach with respect to MC is further accompanied by increased or comparable accuracy, thereby providing compelling evidence for the usefulness of gPC in the study of models in neuroscience. When the significant acceleration of the computation is accompanied by loss of accuracy, the method is still precious to identify regions of interest in the uncertain parameter space, where the resolution can be further refined by resorting to multi-element gPC, see e.g. \cite{wan2006multi}; this is an interesting direction for future research.

We have considered models that range from single neuron to interconnected neurons. A first case study of robustness for the Hindmarsh-Rose model confirms that the tuned collocation method can rapidly identify the robustness region for the plateau bursting regime. The study highlights a transition point to alternative regimes that is consistent with the literature, and further supports the suitability of gPC surrogates to efficiently investigate complex models of biological interest. Future studies may expand the robustness analysis to all HR regimes and to other models, thus generating new insights into their robustness properties.




\section*{Acknowledgments}
This work was funded by the European Union through the ERC INSPIRE grant (project number 101076926). Views and opinions expressed are however those of the authors only and do not necessarily reflect those of the European Union or the European Research Council Executive Agency. Neither the European Union nor the European Research Council Executive Agency can be held responsible for them.

\bibliography{sample.bib}

\end{document}